\documentclass[a4paper,11pt]{article}
\usepackage[top=1.8cm,bottom=1.8cm,left=1.8cm,right=1.8cm,asymmetric]{geometry}
\usepackage{url}
\usepackage{paralist}
\usepackage{authblk}
\usepackage[colorlinks=true,hyperfootnotes=true]{hyperref}

\title{Top Tips to Make Your Research Irreproducible}

\author[1]{Neil P. Chue Hong}
\author[2]{Tom Crick}
\author[3]{Ian P. Gent}
\author[4]{Lars Kotthoff}
\author[5]{Kenji Takeda}
\affil[1]{Software Sustainability Institute, University of Edinburgh, UK}
\affil[2]{Department of Computing \& Information Systems, Cardiff Metropolitan University, UK}
\affil[3]{School of Computer Science, University of St Andrews, UK}
\affil[4]{Insight Centre for Data Analytics, University College Cork, Ireland}
\affil[5]{Microsoft Research, Cambridge, UK}
\affil[1]{\url{n.chuehong@software.ac.uk} ~~~~ \protect\url{http://www2.epcc.ed.ac.uk/\~neilc/} }
\affil[2]{\url{tcrick@cardiffmet.ac.uk} ~~~~ \protect\url{http://drtomcrick.com/}}
\affil[3]{\url{ian.gent@st-andrews.ac.uk} ~~~~ \protect\url{http://ian.gent}}
\affil[4]{\url{lars.kotthoff@insight-centre.org} ~~~~ \protect\url{http://4c.ucc.ie/\~larsko/}}
\affil[5]{\url{kenji.takeda@microsoft.com} ~~~~ \protect\url{http://research.microsoft.com/en-us/people/kenjitak/}}

\date{1 April 2015}

\begin{document}
\maketitle

We have noticed (and contributed to) a number of manifestos, guides
and top tips on how to make research
reproducible~\cite{prlic+procter:2012,sandve-et-al:2013,gent:2013,joppa-et-al:2013,gent_recomputation.org_2014,wilson-et-al:2014,goble:2014,crick-et-al_wssspe2,crick-et-al_recomp2014,stodden+miguez:2014};
however, we have seen very little published on how to make research
\emph{irreproducible}.

Irreproducibility is the default setting for all of science, and
irreproducible research is particularly common across the
computational sciences. The study of making your work irreproducible
without reviewers complaining is a much neglected area; we feel
therefore that by encapsulating some of our top
tips\footnote{{\emph{N.B.}} We are by no means claiming this is an
exhaustive list for making your research irreproducible...} on
irreproducibility, we will be filling a much-needed gap in the domain
literature. By following our starter tips, you can ensure that if your
work is wrong, nobody will be able to check it; if it is correct, you
will make everyone else do disproportionately more work than you to
build upon it. In either case you are the beneficiary.

It is an unfortunate convention of science that research should
pretend to be reproducible; our top tips will help you salve the
conscience of certain reviewers still bound by this fussy
conventionality, enabling them to enthusiastically recommend
acceptance of your irreproducible work.

\begin{enumerate}
\item {\textbf{Think ``Big Picture''.}} People are interested in the
science, not the dull experimental setup, so don't describe it. If
necessary, camouflage this absence with brief, high-level details of
insignificant aspects of your methodology.
\item {\textbf{Be abstract.}} Pseudo-code is a great way of
communicating ideas quickly and clearly while giving readers no chance
to understand the subtle implementation details (particularly the
custom toolchains and manual interventions) that actually make it work.
\item {\textbf{Short and sweet.}} Any limitations of your
methods or proofs will be obvious to the careful reader, so there is
no need to waste space on making them explicit\footnote{Space saved in
this way can be used to cite the critical papers in the field,
i.e. those papers that will inflate your own (as well as potential
reviewers') h-index.}. However much work it takes colleagues to fill
in the gaps, you will still get the credit if you just say you have
amazing experiments or proofs (with a hat-tip to Pierre de Fermat:
``{\emph{Cuius rei demonstrationem mirabilem sane detexi hanc marginis
exiguitas non caperet.}}'').
\item {\textbf{The deficit model.}} You're \emph{the} expert in the
domain, only you can define what algorithms and data to run
experiments with. In the unhappy circumstance that your methods do not
do well on community curated benchmarks, you should create your own
bespoke benchmarks and use those (and preferably not make them
available to others).
\item {\textbf{Don't share.}} Doing so only makes it easier for other
people to scoop your research ideas, understand how your code actually
works\footnote{An exemplary example:
\url{http://www.phdcomics.com/comics.php?f=1689}} instead of why you
say it does, or worst of all to understand that your code doesn't
actually work at all.
\end{enumerate}

However, our most important tip is deceptively but beautifully simple:
{\textbf{to ensure your work is irreproducible, make sure that you
cannot reproduce it yourself}}. If you were able to reproduce it,
there would always be the danger of somebody else being able to do
exactly the same as you. Much else follows from this; for example,
complete confidence in your own inability to reproduce work saves
tedious time revising your work on advice from reviewers: if you are
unable to browbeat the editor into accepting it as is, you can always
resubmit elsewhere. A major advantage of this key insight is that no
strict discipline is required to ensure self-irreproducibility: in our
experience, irreproducibility can happily occur after only the tiniest
amount of carelessness at one of any number of stages.

We make a simple conjecture: {\textbf{an experiment that is
irreproducible is exactly equivalent to an experiment that was never
carried out at all}}. The happy consequences of this conjecture for
experts in irreproducibility will be published elsewhere, with
extremely impressive experimental support.\\

\noindent We close with a mantra for scientists interested in irreproducibility:
\begin{quote}
\emph{After
Publishing
Research,
Irreproducibility 
Lets
False
Observations
Obtain
Longevity!}
\end{quote}










\bibliographystyle{unsrt}
\bibliography{ITT}

\end{document}